\documentclass[12pt]{article}
\usepackage{epsf}
\usepackage{graphicx}

\begin{document}
\title
{A non-universal transition to asymptotic freedom in low-energy
quantum gravity}
\author
{Michael A. Ivanov \\
Physics Dept.,\\
Belarus State University of Informatics and Radioelectronics, \\
6 P. Brovka Street,  BY 220027, Minsk, Republic of Belarus.\\
E-mail:  michai@mail.by.}

\maketitle

\begin{abstract} The model of low-energy quantum gravity by the
author has the property of asymptotic freedom at very short
distances. The character of transition to asymptotic freedom is
studied here. It is shown that this transition is not universal,
but the one obeys the scaling rule: the range of this transition
in units of $r/E^{1/2}$, where $r$ is a distance between particles
and $E$ is an energy of the screening particle, is the same for
any micro-particle. This range for a proton is between $10^{-11} -
10^{-13}$ meter, while for an electron it is approximately
between $10^{-13} - 10^{-15}$ meter.
\end{abstract}
\section[1]{Introduction }
Recently, it was shown by the author \cite{501} that asymptotic
freedom appears at very short distances in the model of low-energy
quantum gravity \cite{500}. In this case, the screened portion of
gravitons tends to the fixed value of $1/2$, that leads to the
very small limit acceleration of the order of $10^{-13} \ m/s^{2}$
of any screened micro-particle. While asymptotic freedom of strong
interactions \cite{20,21} is due to the anti-screening effect of
gluons, the gravitational one is caused by the external character
of graviton flux and the limited rise of the screened portion at
ultra short distances. In this paper, I consider how a transition
occurs from the inverse square law to almost full asymptotic
freedom. The most important property of this transition is its
non-universality: for different particles it takes place in
different distance ranges, and the order of these ranges is
terribly far from the Planck scale where one usually waits of
manifestations of quantum gravity effects.

\section[2]{The screened portion of gravitons at very short
distances}
\begin{figure}[th]
\epsfxsize=9.0cm \centerline{\epsfbox{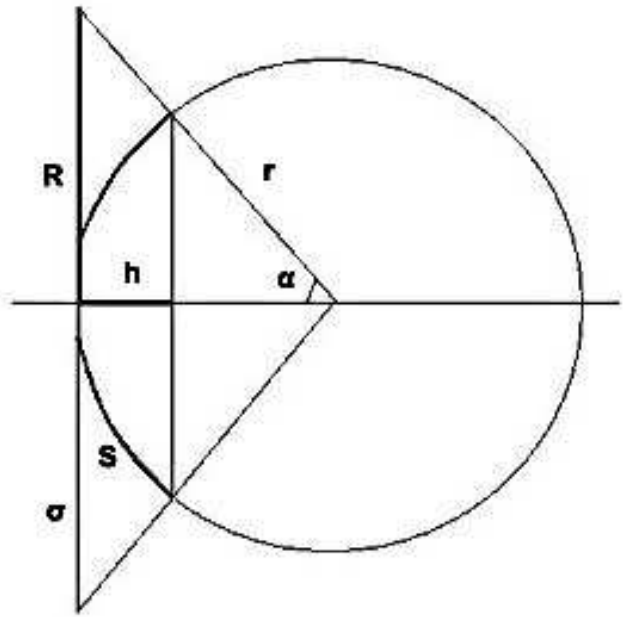}} \caption{To the
computation of the screened portion of gravitons at small
distances: $\sigma$ is the cross-section, $S$ is a square of the
spherical segment of a hight $h$. }
\end{figure}
In the model of low-energy quantum gravity \cite{500}, the
condition of big distances:
\begin{equation}
\sigma (E_{2},<\epsilon_{2}>) \ll 4 \pi r^{2}.
\end{equation}
should be accepted to have the Newton law of gravitation. I use
here the notations of \cite{500}: $ \sigma (E_{2},<\epsilon_{2}>)$
is the cross-section of interaction of graviton pairs with an
average pair energy $<\epsilon_{2}>$ with a particle having an
\begin{figure}[th]
\epsfxsize=9.0cm \centerline{\epsfbox{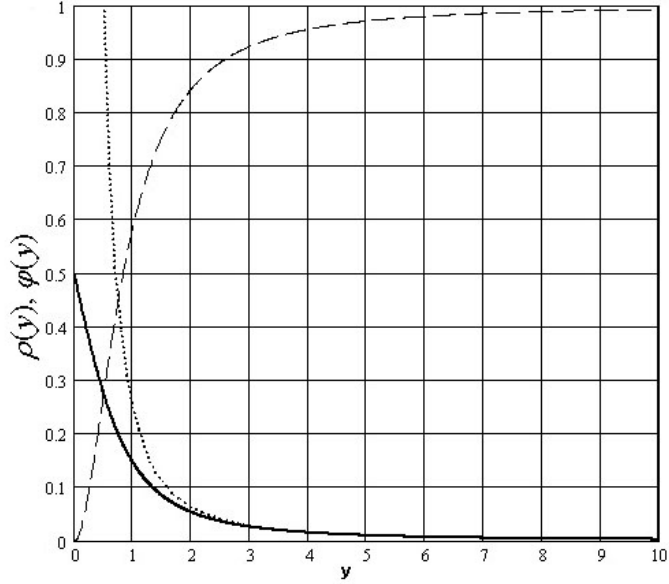}} \caption{Graphs of
the functions $\rho(y) \ (solid), \ \rho(y)_{cl} \ (dot), \
\varphi(y) \ (dash) $ . }
\end{figure}
energy $E_{2}$, $r$ is a distance between particles 1 and 2. As it
was shown in \cite{502}, the equivalence principle should be
broken at distances $\sim 10^{-11} \ m,$ when the condition (1) is
violated for a proton-mass particle. The ratio
\begin{equation}
\sigma (E_{2},<\epsilon_{2}>) / 4 \pi r^{2}.
\end{equation}
describes the screened portion of gravitons for a big distance
$r$. For small $r$, let us consider Fig. 1, where $R=(\sigma
(E_{2},<\epsilon_{2}>)/\pi)^{1/2}$, $S$ is the screening area (the
square of the spherical segment of the hight $h$), and $\alpha$ is
an angle for which $\cot \alpha=r/R \equiv y$. Then we get for
$S:$
\begin{equation}
S=2\pi r^{2}(1- y/(1+y^{2})^{1/2}),
\end{equation}
and it is necessary to replace the ratio (2) by the following one:
\begin{equation}
\rho(y) \equiv S / 4 \pi r^{2}= (1- y/(1+y^{2})^{1/2})/2.
\end{equation}
I rewrite (2) as $\rho(y)_{cl} \equiv 1/4y^{2},$ and I introduce
the ratio of these functions:
\begin{equation}
\varphi(y) \equiv
\rho(y)/\rho(y)_{cl}=2y^{2}(1-y/(1+y^{2})^{1/2}).
\end{equation}
In Fig. 2, the behavior of the functions $\rho(y), \ \rho(y)_{cl},
\ \varphi(y)$ is shown. The upper limit of $\rho(y)$ by
$y\rightarrow 0$ is equal to 1/2; namely this property of the
function leads to asymptotic freedom \cite{501}.
\par
\begin{figure}[th]
\epsfxsize=9.0cm \centerline{\epsfbox{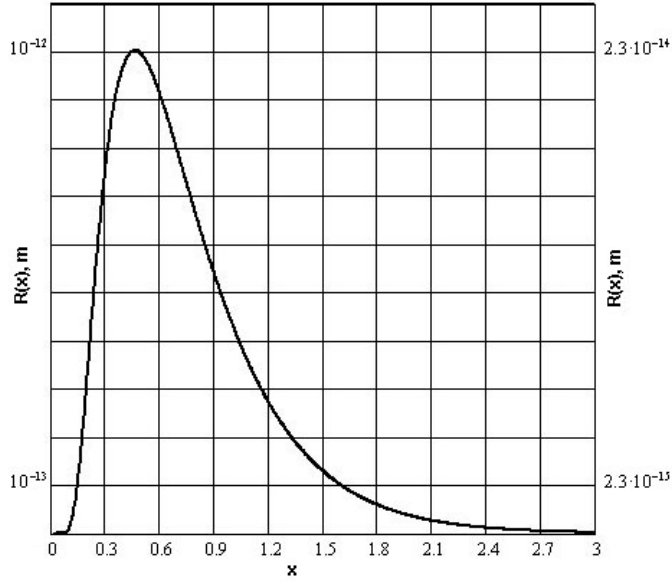}} \caption{The
function $R(x)$ for the two cases: $E_{2}=m_{p}c^{2}$ (the left
logarithmic vertical scale) and $E_{2}=m_{e}c^{2}$ (the right
logarithmic vertical scale).}
\end{figure}
In this model, the cross-section $\sigma (E_{2},<\epsilon_{2}>)$
is equal to \cite{500}:
\begin{equation}
\sigma (E_{2},<\epsilon_{2}>)={ D kT E_{2}2x
(1-\exp(-(\exp(2x)-1)^{-1}))(\exp(2x)-1)^{-2} \over
\exp((\exp(2x)-1)^{-1}) \exp((\exp(x)-1)^{-1})},
\end{equation}
where $T=2.7K$ is the temperature of the graviton background,
$x\equiv \hbar\omega/kT$, $\hbar\omega$ is a graviton energy, the
new constant $D$ has the value: $D=0.795 \cdot 10^{-27}{m^{2} /
eV^{2}}.$ The quantity $R(x)$ has been computed for the two cases
(see Fig. 3): $E_{2}=m_{p}c^{2}$ (the left vertical axis on Fig.
3) and $E_{2}=m_{e}c^{2}$ (the right vertical axis on Fig. 3),
where $m_{p}$ and $m_{e}$ are masses of a proton and of an
electron, correspondingly.
\section[3]{A non-universal transition to asymptotic freedom}
To find the net force of gravitation $F=F_{2}/2$ at a small
distance $r$, we should replace the factor $\sigma
(E_{2},<\epsilon_{2}>) / 4 \pi r^{2}$ in Eq. (31) of \cite{500}
with the more exact factor $S / 4 \pi r^{2}$. Then we get:
\begin{equation}
F(r)= {4 \over 3} \cdot {D(kT)^{5}E_{1} \over
\pi^{2}\hbar^{3}c^{3}} \cdot g(r),
\end{equation}
where $E_{1}$ is an energy of particle 1, and  $g(r)$ is the
function of $r$:
\begin{equation}
g(r) \equiv \int_{0}^{\infty}{ x^{4}
(1-\exp(-(\exp(2x)-1)^{-1}))(\exp(2x)-1)^{-3} \over
\exp((\exp(2x)-1)^{-1}) \exp((\exp(x)-1)^{-1})} \cdot \rho(y) d x,
\end{equation}
where $y=y(r,x)=r/R(x)$. By $r\rightarrow 0$, this function's
\begin{figure}[th]
\epsfxsize=9.0cm \centerline{\epsfbox{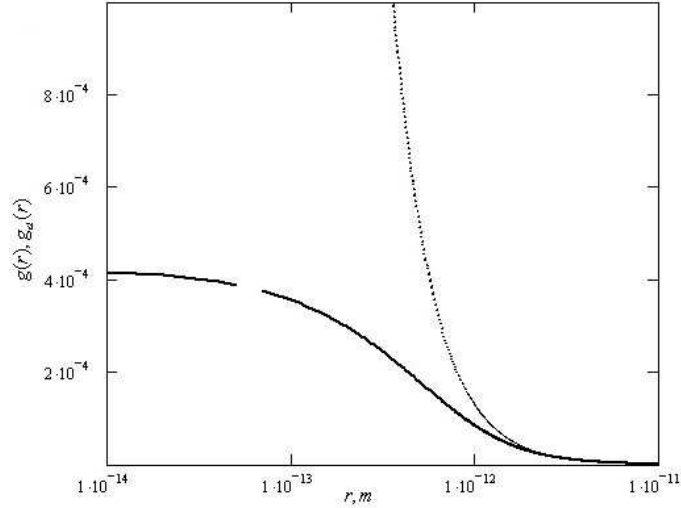}} \caption{Graphs of
the functions $g(r)$ (solid) and $g_{cl}(r)$ (dot) for the case
$E_{2}=m_{p}c^{2}$. }
\end{figure}
limit for any $E_{2}$ is: $g(r)\rightarrow I_{5}= 4.24656 \cdot
10^{-4}$ \cite{501}. Because breaking the inverse square law is
described with this new function, it will be convenient to
introduce the function $g_{cl}(r) \propto 1/r^{2}$ which differs
from $g(r)$ only with the replacement: $\rho(y)\rightarrow
\rho_{cl}(y)$. Graphs of these two functions, $g(r)$ and
$g_{cl}(r)$, are shown in Fig. 4 for the case $E_{2}=m_{p}c^{2}$.
For comparison, graphs of the function $g(r)$ are shown in Fig. 5
for
\begin{figure}[th]
\epsfxsize=9.0cm \centerline{\epsfbox{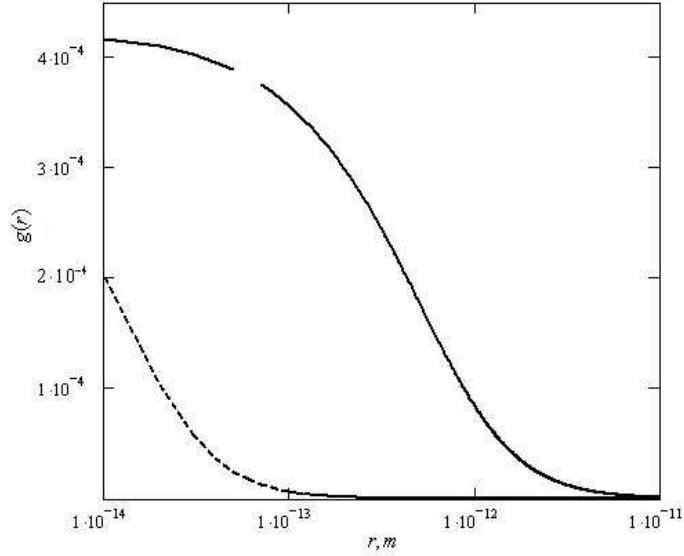}} \caption{Different
transition to the limit value of the function $g(r)$ by
$E_{2}=m_{p}c^{2}$ (solid) and by $E_{2}=m_{e}c^{2}$ (dot).}
\end{figure}
the following different energies: $E_{2}=m_{p}c^{2}$ and
$E_{2}=m_{e}c^{2}$. The functions have the same limit by
$r\rightarrow 0$, but the most interesting thing is their
different transition to this limit when $r$ decreases.
\begin{figure}[th]
\epsfxsize=9.0cm \centerline{\epsfbox{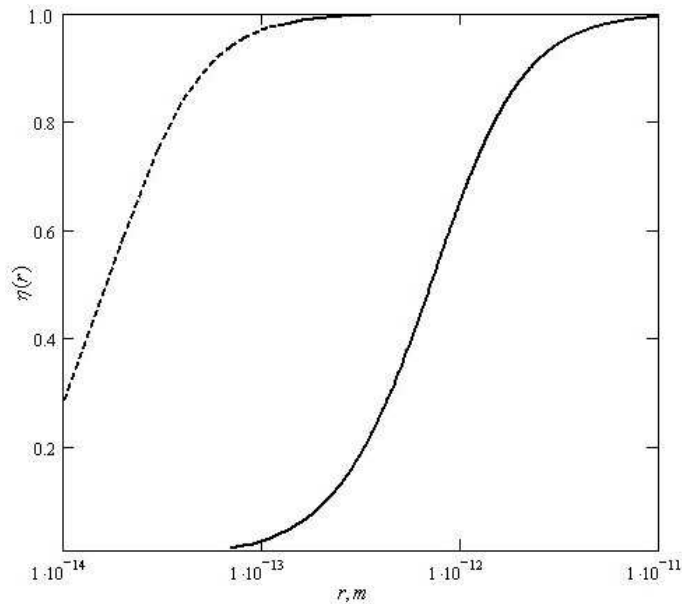}} \caption{A
non-universal transition to the limit value of unity of the
function $\eta(r)$ by $E_{2}=m_{p}c^{2}$ (solid) and by
$E_{2}=m_{e}c^{2}$ (dot). }
\end{figure}
\par
To underline this non-universal behavior, we can compute the
ratio:
\begin{equation}
\eta(r) \equiv g(r)/g_{cl}(r),
\end{equation}
which aims to unity by big $r$. Graphs of this function $\eta(r)$
are shown in Fig. 6 for the same energies: $E_{2}=m_{p}c^{2}$ and
$E_{2}=m_{e}c^{2}$. As we see in this picture, the range of
transition for a proton is between $10^{-11} - 10^{-13}$ meter,
while for an electron it is between $10^{-13} - 10^{-15}$ meter.
So as $y(r,x)=r/R(x)\propto r/E_{2}^{0.5}$, it is obvious that the
functions $g(r/E_{2}^{0.5})$ and $\eta(r/E_{2}^{0.5})$ are
universal for any energy $E_{2}$ of a micro-particle. This scaling
law means, for example, that if deviations from the inverse square
law begins for a proton at $r_{0}\sim 10^{-11} \ m$, then for a
particle with an energy of $E_{2x}$, the same deviations appear at
$r_{0x}=r_{0}\cdot (E_{2x} / m_{p}c^{2})^{0.5}$.
\section[5]{Conclusion}
The considered model has the two unexpected properties: asymptotic
freedom and a non-universal transition to it. As distinct from
QCD, at very small distances the attractive force of gravitation
doesn't decrease when $r\rightarrow 0$, instead, it remains only
finite and very small - but its limit value is the maximal
possible one. Perhaps, it would be better to say that gravity
between micro-particles gets a saturation at short distances in
this model.

\end{document}